\documentstyle[11pt]{article}

\begin{document}

\author{{\bf J. Acacio de Barros } \\
{\em Departamento de F\'{\i}sica -- ICE}\\
{\em Universidade Federal de Juiz de Fora}\\
{\em 36036-330 Juiz de Fora MG Brazil} \and {\bf N. Pinto-Neto} \\
{\em CBPF -- Lafex}\\
{\em Rua Dr. Xavier Sigaud 150 -- Urca}\\
{\em 22290-220 Rio de Janeiro RJ Brazil}}
\title{Comments on the Quantum Potential Approach to a Class of Quantum
Cosmological Models}
\date{}
\maketitle

\begin{abstract}
In this comment we bring attention to the fact that when we apply the
ontological interpretation of quantum mechanics, we must be sure to use it
in the coordinate representation. This is particularly important when
canonical tranformations that mix momenta and coordinates are present.
This implies that some of the results obtained by A. B\l aut and
J. Kowalski-Glikman are incorrect.
\end{abstract}

\section{Introduction}

In a recent paper, A. B\l aut and J. Kowalski-Glikman \cite{BlautGlikman} tried to interpret a classe of quantum cosmological models in terms of
Bohm's causal interpretation of quantum mechanics \cite{Bohm52}.
Following a formalism developed by Ashtekar {\em et al}. \cite{Ashtekar},
they applied a canonical quantization
procedure to a restricted class of spacetimes, 
whose Hamiltonian
constraint has been put in a simple form after a non-trivial canonical
transformation. Then,
with the standard decomposition of the wavefunction in polar form, they
obtained from the Wheeler-DeWitt equation a modified Hamilton-Jacobi
equation with an extra quantum potential term. 
From the solutions of this equation, they computed 
Bohmian trajectories, and obtained possible scenarios for the universe
modeled by the given wavefunction. In this comment, we will show that the
interpretation presented in reference \cite{BlautGlikman}
is not adequate because when using Bohm's interpretation we
have to make sure that the wavefunction we use is in the coordinate
representation.

\section{The Classical Model}

The minisuperspace examples A. B\l aut and J. Kowalski-Glikman 
\cite{BlautGlikman} used were
classes of diagonal, spatially homogeneous cosmological models which admit
intrinsic, multiply transitive symmetry groups (DIMT models). The spatially homogeneous
diagonal 4-metric can be expressed in the form 
\begin{equation}
ds^2=-N^2(t)\,dt^2+\sum_{i=1}^3 \exp (2\beta ^i) (\omega ^i)^2  \label{diagonalmetric}
\end{equation}
where $N(t)$ is the lapse function, and $\omega ^i$ is a basis of spatial
1-forms which are left invariant by the action of the isometry group. 

Since in their
paper, A. B\l aut and J. Kowalski-Glikman analyzed the
case of a plane wave in a Bianchi type IX spacetime, we will focus our
attention on this case.
Imposing the Taub gauge $N_t=12\exp (3\beta ^0)$, we can express the scalar
Hamiltonian constraint for the Bianchi IX model as 
\begin{equation}
{\cal H=H}_0+{\cal H}_{+},  \label{hamiltdecomp}
\end{equation}
where 
\begin{eqnarray}
{\cal H}_0 &=&-\frac 12\bar{p}_0^2 - 24 \exp (2\sqrt{3}\bar{\beta}^0),
\label{Hconstraint} \\
{\cal H}_{+} &=&\frac 12\bar{p}_{+}^2+6\exp (-4\sqrt{3}\bar{\beta}^{+}).
\label{Lastconstraint}
\end{eqnarray}
In the above expressions we have 
\begin{equation}
\left( {\beta}^1,{\beta}^{2},{\beta}^{3}\right) =\sqrt{3}
\left( \bar{\beta} ^0+\bar{\beta} ^{+},\bar{\beta} ^0+\bar{\beta}^{+}, -\bar{\beta} ^{+}\right).  
\label{misner}
\end{equation}

The separable form of the scalar constraint presented above makes it
possible to perform a canonical transformation that simplify its form.
It is given by 
\begin{equation}
\tilde{p}_A=\sqrt{\bar{p}^2_A+a\exp (2b\bar{\beta}^A}) ,
\label{ptilde}
\end{equation}
and 
\begin{equation}
\tilde{\beta}=\frac 1b\left[ \log \left( -\bar{p}+\sqrt{p^2+a\exp (2b\bar{%
\beta})}\right) -\log \left( \sqrt{a}\exp (b\bar{\beta})\right) \right] .
\label{transformation}
\end{equation}
where $a$ and $b$ can be read from equtions (\ref{Hconstraint}) and
(\ref{Lastconstraint}).
With the transformations (\ref{ptilde}) and (\ref{transformation}), 
the Hamiltonian constraint (\ref{hamiltdecomp}) for the Bianchi IX model
acquires the simple form 
\begin{equation}
{\cal H}= -\frac 12\left( \tilde{p}_0^2-\tilde{p}_{+}^2\right) =0.
\label{kleingordon}
\end{equation}

\section{Bohm's Trajectories}

Let us start with the equation 
\begin{equation}
\hat{{\cal H}}\psi (\bar{\beta}^A)=\left[
\frac 12 {\Box}+{\cal V}(\bar{\beta}^A)\right]\psi (\bar{\beta}^A)=0  
\label{Schroedinger}
\end{equation}
where ${\Box}\equiv \eta ^{AB}\partial /\partial \bar{\beta}^A\partial
/\partial \bar{\beta}^B$ with $\eta ^{AB} = {\rm diag} (1,-1)$ and
\begin{equation}
\label{potential}
{\cal V}(\bar{\beta}^A)=- 24 \exp (2\sqrt{3}\bar{\beta}^0)+6\exp (-4\sqrt{3}\bar{\beta}^{+}).
\end{equation} 
Equation (\ref{Schroedinger}) is the the Wheeler-de Witt equation of the
Bianchi IX model coming from the Hamiltonian constraint (\ref{hamiltdecomp}). 
We will use the standard polar decomposition for the wavefunction 
\begin{equation}
\psi =R(\bar{\beta}^A)\exp (\frac i\hbar S(\bar{\beta}^A)).  
\label{PolarForm}
\end{equation}
Substitution of $\psi $ in the form (\ref{PolarForm}) into (\ref
{Schroedinger}) results in 
\[
-\frac 12\eta ^{AB}\frac{\partial S}{\partial \bar{\beta}^A}\frac{\partial S}{
\partial \bar{\beta}^B}+{\cal V}-
\frac{\hbar ^2}2\frac 1R {\Box}R=0, 
\]
which can be seen as a Hamilton-Jacobi like equation plus a quantum
potential term. Adopting the Bohm's interpretation, one can {\em postulate} the momenta as 
\begin{equation}
\bar{p}_A = \frac{\partial S}{\partial \bar{\beta}^A}=
\eta _{AB} \frac{{\rm d}\bar{\beta}^B}{{\rm d}t},  
\label{Trajectory}
\end{equation}
where the parameter $t,$ the time, was introduced. The trajectories followed
by the system are solutions of the equation (\ref{Trajectory}), and are
different from the classical ones due to the quantum potential.

In the tilde variables, the Wheeler-DeWitt equation becomes (see equation
(\ref{kleingordon})               )
\begin{equation}
\frac 12 {\Box}^{\prime} \psi =0,
\label{qkg}
\end{equation}
where $\frac 12 {\Box}^{\prime} \equiv \eta ^{AB}\partial /\partial \tilde{\beta}^A\partial
/\tilde{\beta}^B$, which is evidently much more simple to solve
then equation (\ref{Schroedinger}).

\section{The Plane Wave Example}

In this Section we will analyze the example given by B\l aut and
Kowalski-Glikman. The wavefunction they interpreted is given by 
\begin{equation}
\psi (\tilde{\beta}_0,\tilde{\beta}_{+})=\exp [i(k+l)U]+\exp [i(k-l)V],
\label{planewave}
\end{equation}
where $U=\tilde{\beta}_0+\tilde{\beta}_{+}$, 
$V=\tilde{\beta}_0-\tilde{\beta}_{+}$, and $k$ and $l$ are real constants. 
It is a solution of the Wheeler-De Witt equation (\ref{qkg})
in the tilde variables.
They obtained Bohmian trajectories by using equation (\ref{Trajectory})
in the tilde variables
\begin{equation}
\tilde{p}_A=\frac{\partial S}{\partial \tilde{\beta}^A}=
\eta _{AB} \frac{{\rm d}\tilde{\beta}^B}{{\rm d}t},  
\label{Trajectory2}
\end{equation}
where $S$ is the phase of the wave function (\ref{planewave}). 
Our main point is that the tilde variables were
obtained from the barred variables by a non-trivial canonical transformation
which mix momenta with coordinates, and hence one cannot apply directly
the Bohm interpretation to these variables by using equation
(\ref{Trajectory2}).

Quantum mechanically, to look for a
canonical transformation means to look for a unitary transformation that
maps wavefunctions from the original set of variables to the new one. In
other words, we need to find the kernel $\langle \tilde{\beta}|\bar{\beta}%
\rangle $, which was obtained in reference \cite{BlautGlikman}. 
However, the canonical transformation (\ref{ptilde}) and (\ref{transformation}) 
to the tilde variables
mix coordinates with momenta. As is well known, Bohm's
intepretation only makes sense in coordinate representation \cite{bohm2,hol}. 
If we make a
canonical tranformation that mixes momenta and coordinates, we may end up
having misleading results. The only safe way to guarantee the correct interpretation of
the solution (\ref{planewave}) is to go back to the wave function expressed in
the original set of variables by using the
kernel $\langle \tilde{\beta}|\bar{\beta}\rangle $, and then use equation 
(\ref{Trajectory}) to obtain the quantum trajectories. The final result will
in general be different from the one obtained directly from the wave
function  (\ref{planewave}) by using equation (\ref{Trajectory2}) in the
tilde variables and then going back to the barred variables by using the
inverse of the tranformation (\ref{ptilde}) and (\ref{transformation}), as
is done in reference \cite{BlautGlikman}. 

Let us illustrate this point with
a simple example showing how the two procedures can give different Bohmian
trajectories. Take the Hamiltonian for the free particle 
\begin{equation}
H=\frac{p^2}{2m}.
\end{equation}
and let us make a canonical transformation to a new set of variables given
by 
\begin{eqnarray}
X &=&\frac{a^2}{\hbar}p+x,  \label{canonica1} \\
P &=&p.  \label{canonica2}
\end{eqnarray}
where $a$ is some constant with dimension of length.
Then, in the new set of variables we have 
\[
H=\frac{P^2}{2m}.
\]
We want to find a kernel that transforms the wavefunction from the original
set of variables to the new one. This is accomplished by solving the
following set of equations:
\begin{eqnarray}
\hat{p}\psi (x) &=&\int_{-\infty }^\infty {\rm d}X\ \langle x|X\rangle 
\hat{P}\psi (X) \\
\hat{x}\psi (x) &=&\int_{-\infty }^\infty {\rm d}X\ \langle x|X\rangle 
(\hat{X}-\hat{P})\psi(X).
\end{eqnarray}
We can easily solve these equations and obtain that 
\begin{equation}
\langle x|X\rangle =e^{-\frac{i}{2a^2}(x-X)^2}.
\end{equation}
Now we can look for a particular solution of the free particle Schroedinger
equation. In the $X,$ $P$ coordinates, one possible solution is the gaussian 
\[
\psi (X,t)=b(t)\exp \left[ -X^2 \left(\frac{a^2-ic(t)}{a^4+c(t)^2}
\right) \right] ,
\]
where $b(t)=\{2/[\pi a^2(1+ic(t)/a^2)^2]\}^{1/4},$ and $
c(t)=2\hbar t/m.$ If we set, to simplify the computations, $2\hbar =m=a^2=1,$
we obtain from 
\[
P=\dot{X}=\frac{\partial S}{\partial X}
\]
the result 
\[
X(t)=\beta (1+t^2)^{1/2},
\]
where $\beta $ is an integration constant. Using now equations (\ref
{canonica1}) and (\ref{canonica2}) we obtain 
\begin{equation}
x(t)=X(t)-2P(t)=\frac{\beta (t-1)^2}{(t^2+1)^{1/2}}.  \label{outra}
\end{equation}

On the other hand, if we make the transformation 
\[
\psi (x,t)=b(t)\int \exp \left[ -\frac{i}{2a^2}(x-X)^2 \right] \exp \left[ -X^2\left( \frac{a^2-ic(t)}{a^4+c(t)^2}\right) \right] {\rm d}X
\]
we get at once 
\begin{equation}
\label{gauss2}
\psi (x,t)=b^{\prime }(t)\exp \left\{ -x^2 \left[ 
\frac{a^2 - i(c^2(t)-2a^2)}{(c^2(t)-2a^2)^2+a^4} \right] \right\} ,
\end{equation}
where 
$b^{\prime}(t)$ is a function of time which is not important for what follows.
Setting again $2\hbar =m=a^2=1$, we get from the equation 
\[
p=\dot{x}=\frac{\partial S}{\partial x}
\]
the solution 
\begin{equation}
x(t)=\beta ^{\prime }[(t-2)^2 + 1]^{1/2},  \label{outra2}
\end{equation}
where $\beta ^{\prime }$ is a constant of integration, which is
different from the solution (\ref{outra})\footnote{Note that 
solution (\ref{outra2}) is consistent with the gaussian
(\ref{gauss2}) while solution (\ref{outra}) is not (see reference
\cite{hol}, chapter 4).}. Hence, the two methods are
inequivalent. 

In conclusion, we must be very careful when we use the causal interpretation 
because the Bohmian trajectories are not invariant under general canonical transformations.
Knowing this fact, what we have to do in the Bianchi IX example is
to map the wave function in the tilde variables into the correspondent
solution in the barred variables, which
are the configuration variables related to the physical metric, and only
after use the causal interpretation to find the Bohmian trajectories.
Applying directly the causal interpretation to the wave function
in the tilde variables
yields wrong Bohmian trajectories in the barred coordinates.

{\bf Acknowledgments}

Part of this work was done while NPN was a PREVI (Special Program for
Visiting Professor/Researchers) fellow at the Physics Department of the
Federal University at Juiz de Fora (UFJF). We wish to thank FAPEMIG (Minas Gerais
State Sponsoring Agency) for partial financial support for this work. NPN
also thanks CNPq for partial financial support.

\end{document}